\documentclass[12pt,preprint]{aastex}
\usepackage{epsf,psfig}

\usepackage{longtable}

\slugcomment{}
\shorttitle{SDSS Arc Search}
\shortauthors{ }
\begin{document}
\title{A Systematic Search for High Surface Brightness Giant Arcs in a Sloan Digital Sky Survey Cluster Sample}

\author{J. Estrada$^1$, J. Annis$^1$, H.T. Diehl$^1$,
    P. B. Hall$^2$, T. Las$^1$, H. Lin$^1$, M. Makler$^3$,
    K. W. Merritt$^1$, V. Scarpine$^1$, S. Allam$^1$ and D. Tucker$^1$.  }
\affil{ $^1$ Fermi National Accelerator Laboratory, P.O.\ Box 500, Batavia, IL 60510, USA \\
          $^2$ Department of Physics and Astronomy,
                York University, 4700 Keele St.,
                Toronto, Ontario, M3J 1P3, Canada \\
          $^3$ Centro Brasileiro de Pesquisas F\'isicas, Rio de Janeiro, Brazil  }

\begin{abstract}

We present the results of a search for gravitationally-lensed
giant arcs conducted on a
sample of 825 SDSS galaxy clusters. 
Both a visual inspection of the images and an automated search were 
performed and no arcs were found. 
This result is used to set an upper limit on the arc probability per cluster.
We present selection functions for our survey,
in the form of arc detection efficiency curves 
plotted as functions of arc parameters,
both for the visual inspection and the automated search. 
The selection function is such that we are
sensitive only to long, high surface brightness arcs with
$g$-band surface brightness $\mu_g \le 24.8$ and length-to-width ratio
$l/w \ge 10$. 
Our upper limits on the arc probability are compatible with previous arc
searches. Lastly, we report on a serendipitous discovery of a giant
arc in the SDSS data, known inside the SDSS Collaboration as Hall's
arc.

\end{abstract}

\section{Introduction}

Clusters of galaxies are the largest 
gravitationally bound structures in the universe.
It has long been recognized
that they constitute very usefeul probes for cosmology
\citep[see, e.g.,][]{PS,MohrFuture,Voit}, and therefore
understanding them in detail has been a major field of research.
Notable progress in the theoretical understanding of
clusters has been achieved in recent years through N-body and
hydrodynamical simulations. At present, different simulation schemes
converge in reproducing several broad properties of clusters
\citep{StaBarbara,robustness}. Furthermore, several properties of
the simulated clusters are in agreement with observations
\citep[see, e.g.,][]{xrayHydro,clustersHubbleVolume,xrayEvolution}.

The strong gravitational lensing of
background galaxies produced by massive clusters is a
probe of cluster structure. This effect has been
observed in many previous studies, with
the most extreme case being the giant gravitational arcs first seen
by \citet{petrosian} and \citet{discovery2}.
Almost a decade later, spectacular images
of these arcs were provided by the excellent resolution of the
Hubble Space Telescope \citep{HSTSci,HST2218ApJ,SmailHST}.
There are now about a hundred clusters with giant arcs imaged by HST
\citep[for a recent compilation, see][]{HST_sands}. Several arc
searches have been carried out with ground based observations,
both for X-ray selected clusters \citep[see,
e.g.,][]{luppino,Cypriano,Campusano} as well as for optically
identified clusters \citep{zaritsky03,RCS}.

The comparison of the abundance of gravitational arcs with
theoretical predictions has been proposed as a tool to constrain
cosmology \citep{arcs1,arcs2,arcs3}. 
The early results indicated that the number of
strongly-lensed arcs greatly exceeded that expected from
$\Lambda$CDM simulations \citep{bartelman98}. More detailed
studies of the theoretical uncertainties and systematic
effects involved in this analysis seem to indicate that the
observations do, in fact, agree with the theoretical expectations
from the standard cosmology \citep{hennawi,dalal, wambs,
oguri}, although uncertainties may still remain \citep{Li}.

The main goal of the work presented here is not in the extraction of
cosmological information from the statistics of gravitational arcs.
Instead it is both (a) to locate high surface brightness giant arcs
that would prove useful in follow-up imaging and spectroscopy
programs, and more directly (b) to
understand the probability for a galaxy cluster to produce a
giant arc as a function of the cluster mass and redshift. We believe 
this is an important aspect to be considered before cosmology can be
extracted from arc statistics and that it will provide a
significant improvement in our understanding of the mass
distribution inside galaxy clusters.

A defining characteristic of previous arc searches is that they
were done in the most massive clusters.
\citet{luppino} searched
for arcs in 38 X-ray selected clusters (redshifts $0.15 \le z \le 0.82$) with
X-ray luminosity $L_x \ge 2\times10^{44}$ ergs/sec
and found arcs in a high fraction of clusters with $L_x > 10^{45}$ ergs/sec,
but no arcs in clusters with $L_x < 4\times10^{44}$ ergs/sec.
\citet{zaritsky03} searched for arcs in 44 optically selected clusters
($0.5 \le z \le 0.8$)
from a list ranked by surface brightness, finding arcs in two clusters.
\citet{RCS} searched $\approx 900$ optically selected clusters ($0.3\le z \le 1.4$)
ranked by a galaxy overdensity parameter, finding arcs in 8 clusters, with
none at $z < 0.64$.  \citet{HST_sands} used a very different technique,
searching for arcs in the heterogeneous sample of clusters that have
been observed with HST; out of 128 clusters ($0.1 \le z \le 0.78$)
45 had tangential arcs, although it is worth noting
that many of the clusters were targeted by HST precisely because
of known arcs.

It is possible to
compute which of the arcs found in high
quality images would be found in images of lower quality, and we
will perform such a comparison. The results so far suggest
looking for arcs in clusters with high X-ray luminosity or high
redshift. \cite{luppino} found a higher frequency of arcs in high
$L_x$ clusters, and since the bremstrahlung emission in
clusters is proportional to $n_e^2$, this result suggests
that arcs are found preferentially in the highest density
clusters. \cite{RCS} found a higher frequency of arcs in higher
redshift clusters, which suggests either an evolution of cluster
structure in a way not fully understood, or perhaps that
the sources lensed are predominately at higher redshifts, where the
lensing cross section would be much larger for $z \sim 0.7$ clusters 
than for those at $z \sim 0.3$ \citep{ho}.
Observationally, however, it is clear that the dominant
variables for successful arc detection are
limiting surface brightness and, in particular,
seeing.

\begin{figure}
\begin{center}
\plottwo{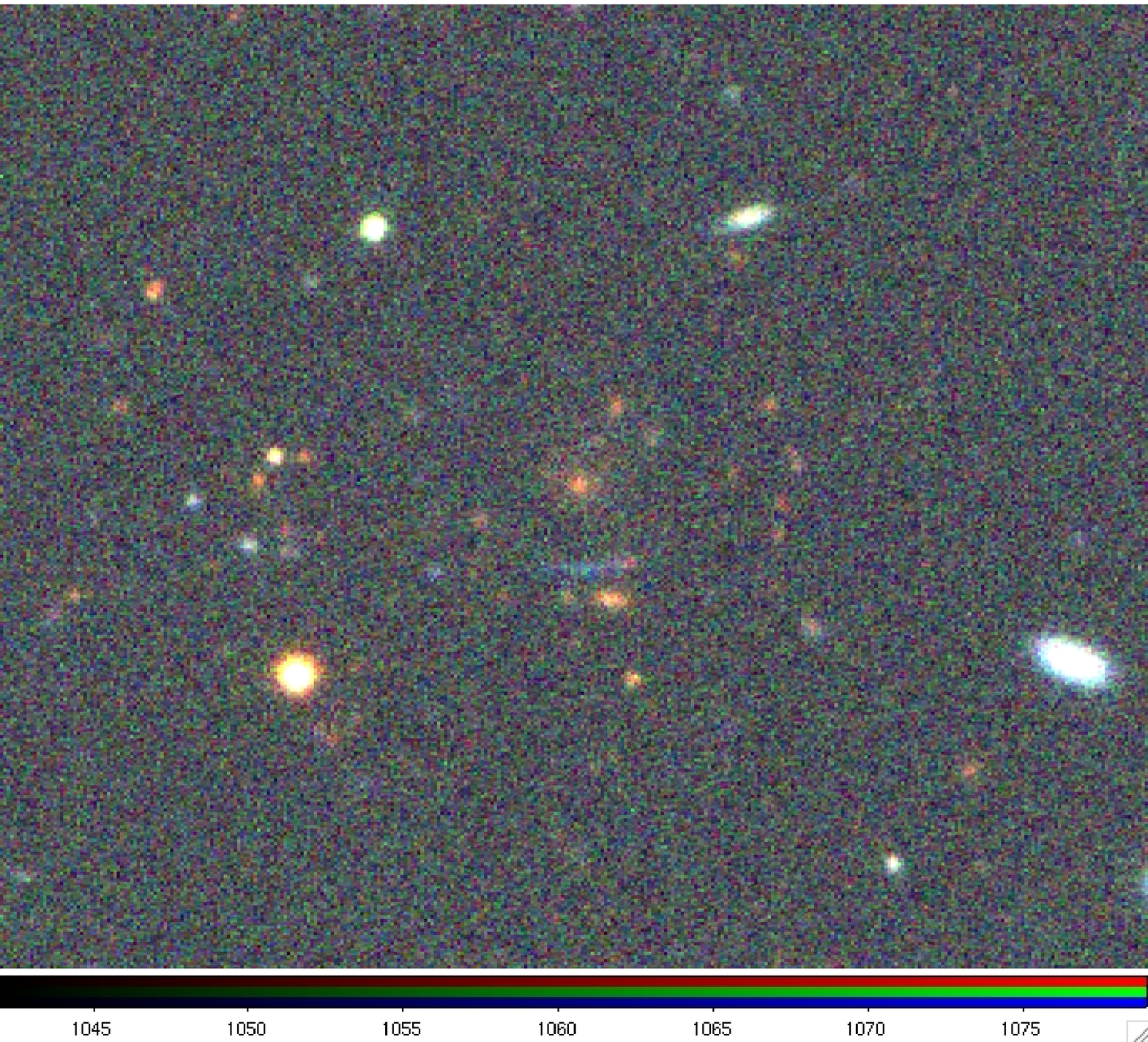}{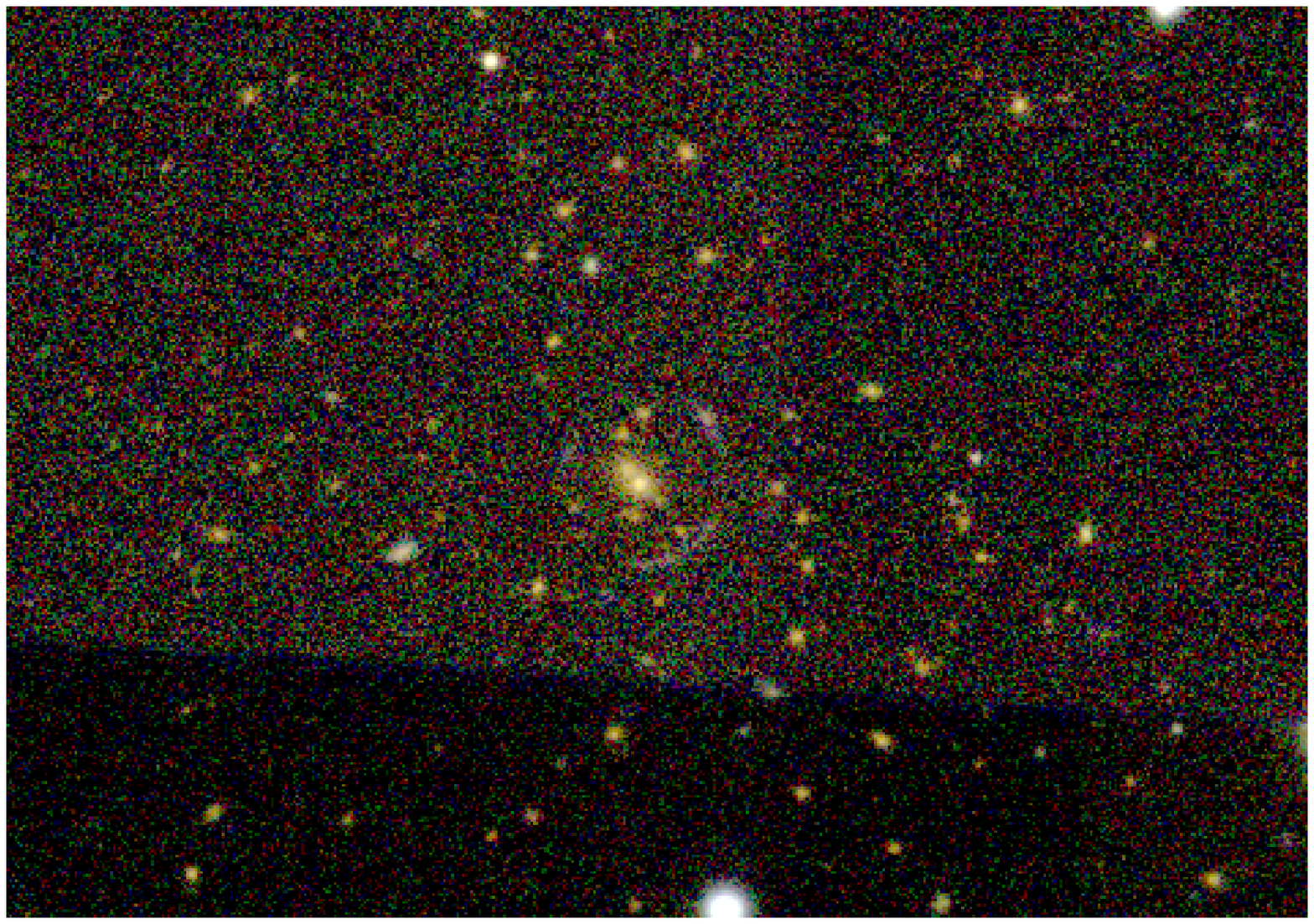}
\end{center}
\caption{On the left, the SDSS image of the RCS1419.2+5361
cluster. The arc was discovered by visual inspection of the RCS
cluster sample \citep{RCS} and is clearly
visible in this color image composed of the SDSS $g$, $r$, and $i$ data. 
On the right, Hall's arc (see Appendix), serendipitously discovered in
the SDSS data. It too is clearly visible in the composite $gri$ image. These
observations, together with the recent discovery of a new arc
system \citep{sahar_arcs}, demonstrate the ability of the SDSS to detect
high surface brightness giant arcs.} \label{RCS_arc}
\end{figure}

The Sloan Digital Sky Survey \citep{York} to date has produced 
multiband $ugriz$ imaging of $8000 $ deg$^2$ of the northern sky, along with
an associated catalog of 1,048,496 spectra, of which 674,749 are
galaxies \citep{DR5}. It has proven useful as a data set for
gravitational lens searches. An example is the multiply-lensed
quasar searches of \cite{pindor} and \cite{oguri06}, which use the
imaging data to select quasar-colored
objects that are larger than a PSF, or groupings of quasar-colored
objects in a small area. A second example of lens searches in the
SDSS data uses the large SDSS spectroscopic data set: \cite{bolton}
searched red galaxy spectra for indications of a second, higher
redshift nebular emission spectrum. This locates strongly-lensed
background galaxies, which are revealed as arcs after applying galaxy
subtraction techniques on HST ACS images. Our program is a search
for strongly-lensed background galaxies behind clusters; that is, a
search for giant arcs. Our ability to locate gravitational arcs in
the SDSS images is demonstrated by recovering arcs observed by other
surveys and by the discovery of new gravitational arcs \citep[see
Fig.~\ref{RCS_arc}, Appendix~\ref{appendix}, and][]{sahar_arcs}.
A similar arc search in a sample of 240 rich SDSS clusters has been
carried out by \cite{Hennawi06} and has found a significant number
of new giant arc systems, but relies on deeper imaging
data to discover the arcs, whereas we use the original 
SDSS imaging data itself.

Our analysis consists of a search for giant gravitational arcs
in a sample of SDSS galaxy clusters sorted by a richness
estimator. Our imaging data does not have the depth or seeing available
to previous searches, but it does have the twin features of homogenity
and large sky coverage. This results in our ability to greatly
increase the number of clusters searched. The search reported
here totals 825 clusters, comparable to the largest previous search.
We intend, in a future paper, to decrease the mass threshold down to
the poor group range in order to search tens or hundreds of
thousands of locations. Furthermore, the redshift range of the
cluster catalog we have searched is predominantly $0.1 \le z \le
0.3$, a regime poorly covered by previous searches.

\clearpage
\clearpage


\section{The Data}

The data set we use is the Sloan Digital Sky Survey (SDSS) 
Data Release 5 \citep[DR5;][]{DR5}. 
The SDSS uses a wide-field drift-scanning mosaic CCD camera
\citep{Gunn98} on a dedicated 2.5m telescope \citep{Gunn06} 
at Apache Point Observatory, New Mexico, to image the sky
in the five photometric bandpasses $ugriz$ \citep{Fukugita96,Stoughton02}.
The SDSS photometric data are processed using pipelines that
address astrometric calibrations \citep{Pier03}, photometric
data reduction \citep{Lupton01}, and photometric calibrations 
\citep{Tucker06,Smith02,Hogg01,Ivezic04}.  Note that the SDSS is 
a survey that trades relatively shallow
exposures -- 55 seconds on a 2.5m telescope, resulting in 
point sources detected at $95\%$ confidence level at $g=22.2$
and $i=21.3$ (AB) -- for a very wide area: 8000 deg$^2$ or $20\%$
of the total sky.

\subsection{The Cluster Sample}

The cluster catalog was constructed using the maxBCG algorithm. This
is a red-sequence locating technique that has shown completeness
levels of $90\%$ at $0.1 \le z \le 0.3$ and down to $N_g = 10$,
where $N_g$ is the number of galaxies on the E/S0 red
sequence brighter than $0.4 L^\star$ and within $1h^{-1}$ Mpc
of the brightest cluster galaxy.  The
correlation of the richness $N_g$ with mass can be
calibrated with independent mass estimators, such as velocity
dispersions or weak lensing shear profiles. A detailed description
of the maxBCG technique is discussed in \cite{SDSS_maxben_alg}. 
\cite{SDSS_maxben_cat} used the SDSS DR5 data to construct
a cluster catalog containing 12,875 rich clusters, plus two orders of
magnitude more clusters of lower masses.
The catalog we used in this work was created using an earlier version of
the maxBCG algorithm \citep[see, e.g.,][]{hansen,WLRASSSDSS}.

One of the defining features of red sequence methods is that
projection effects are present only at a minimal level: if the
clusters differ by $\Delta z < 0.05$ then they can be confused
as a single object. The resulting purity (lack of false positives)
is $\approx 95\%$, and the completeness (lack of false negatives) is
$\approx 85\%$ above $1\times 10^{14}M_{\sun}$, as determined by running
this algorithm on
mock catalogs \citep{SDSS_maxben_cat}.

\begin{figure}
\begin{center}
\plotone{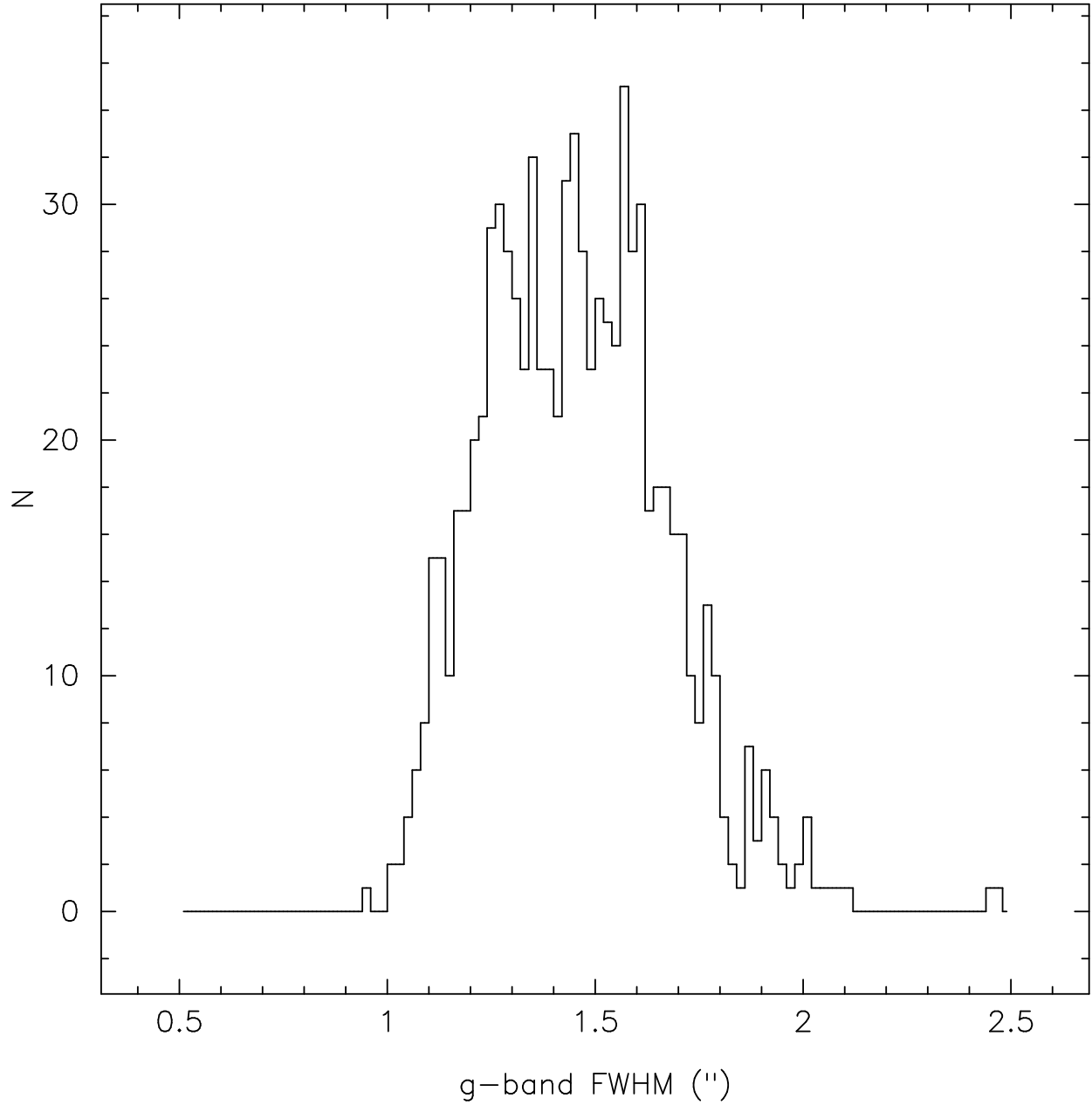}
\end{center}
\caption{The $g$-band seeing distribution for the SDSS images of
the 825 clusters inspected in this paper. The distribution is well
fit by a gaussian of mean 1.44'' and standard deviation 0.20''.}
\label{seeing}
\end{figure}

\begin{figure}
\begin{center}
\plotone{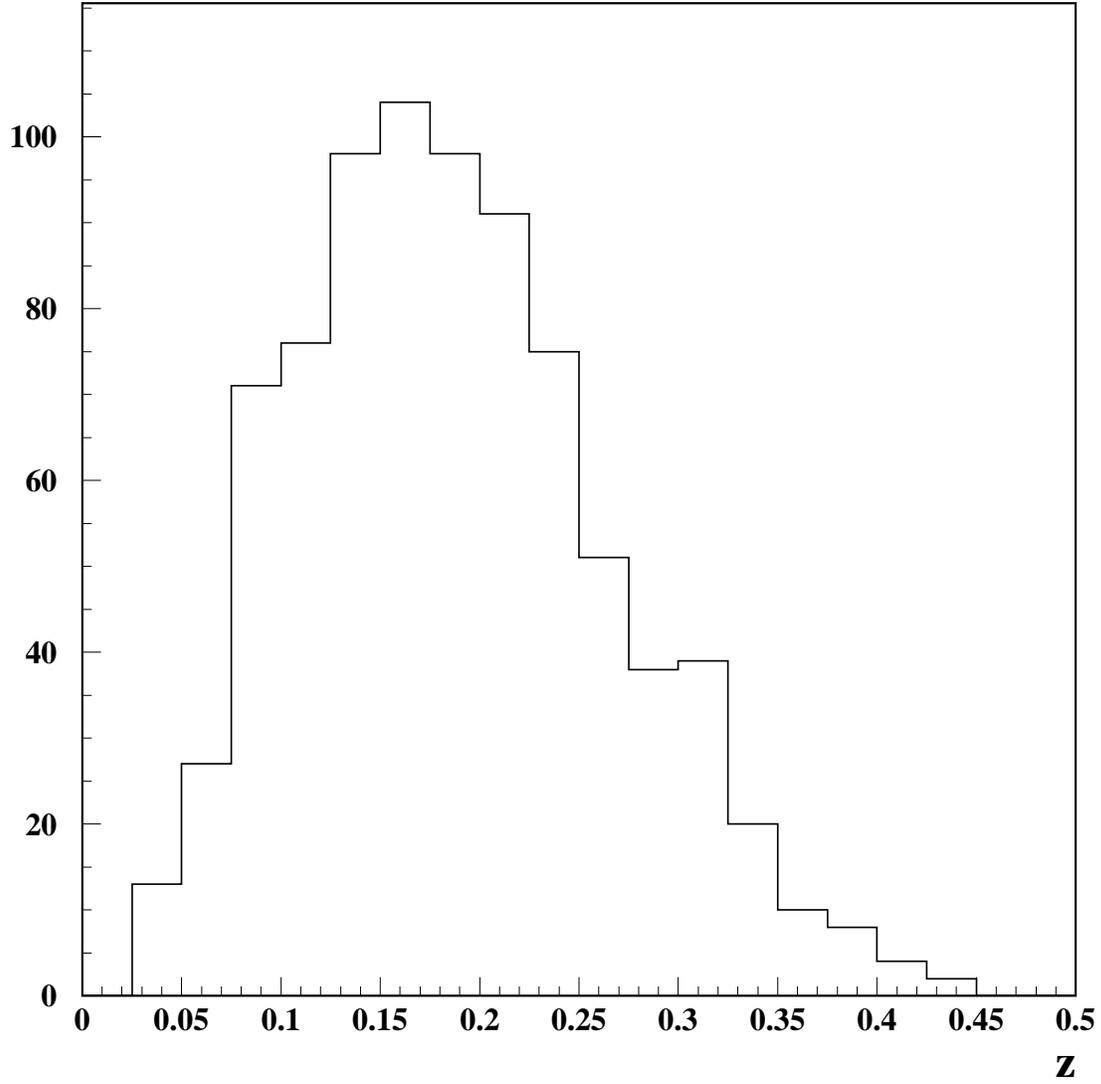}
\end{center}
\caption{The redshift distribution for the 825 clusters
inspected for this paper. We have good statistics in
the redshift range $0.1 < z < 0.3 $.} \label{z} 
\end{figure}

We selected a sample of 825 clusters with $N_g \ge 20$,
approximately 50\% of the $N_g \ge 20$ clusters in our sample.
The whole sample was not searched due to a variety of
technical reasons, but lack of a complete sample is not important
for the results of this paper. The distribution of $g$-band seeing
of the cluster images is shown in Fig.~\ref{seeing} and has a mean
of $1.44''$; this is important because the seeing is the single dominating
factor in our ability to detect arcs. The cluster redshift distribution is
shown in Fig.~\ref{z}. The richest
clusters in our sample are presented in Table~\ref{clusterlist}, and
the table listing our full cluster sample is available electronically.

\clearpage

\clearpage
\clearpage

While we do not use the masses of the clusters directly in this
work, we do have a rough mass calibration.
Stacked cluster spectroscopic velocity dispersions, plus
isothermal sphere fits to stacked cluster weak lensing shear
profiles, both from early work in the SDSS EDR 
\citep[see, e.g.][]{bahcall,WLRASSSDSS}, were combined to produce a
scaling relation between velocity dispersion $\sigma_v$ and $N_g$ given by
$ \sigma_v \approx  85 N_g^{0.6}$ km/s.
Modeling the cluster as a singular isothermal sphere, the
mass enclosed within a radius having an overdensity of 200
with respect to the critical density is \citep[see, e.g.][]{bryan98}
\begin{equation}
M_{200} =  \Big( \frac{ \sigma_v} { \sigma_0}  \Big) ^ 3
\frac{1}{E(z) h} \quad 10^{15} M_{\sun} , \label{mass}
\end{equation}
where $\sigma_0=1148$ km/s, $E(z) = \sqrt{ \Omega_\Lambda +
\Omega_m(1+z)^3+(1-\Omega_\Lambda - \Omega_m)(1+z)^3}$, $h$ is the
Hubble constant in units of 100 km/s/Mpc, and
$\Omega_m$ and $\Omega_\Lambda$ are the energy densities of
cold/presureless normal matter and the cosmological
constant, respectively, in units of the critical density. 
The $\sigma_v$-$N_g$ relation has a
significant scatter. Indirect methods, such as (a) running the
cluster finding algorithm on mock catalogs based on N-body
simulations and (b) measuring the higher moments of the velocity
distributions of galaxies in stacked cluster samples, suggest that
the scatter is $\Delta M_{200}/ M_{200} \sim 40 \%$
at high $N_g$ and increases at lower $N_g$.
A more sophisticated calibration using weak lensing measurements of
the full maxBCG catalog is given in
\cite{SDSS_maxben_mcal} and \cite{WLtheory}. 
In future work we expect to use the
current maxBCG catalog and mass calibrations.

\clearpage
\begin{figure}
\begin{center}
\plotone{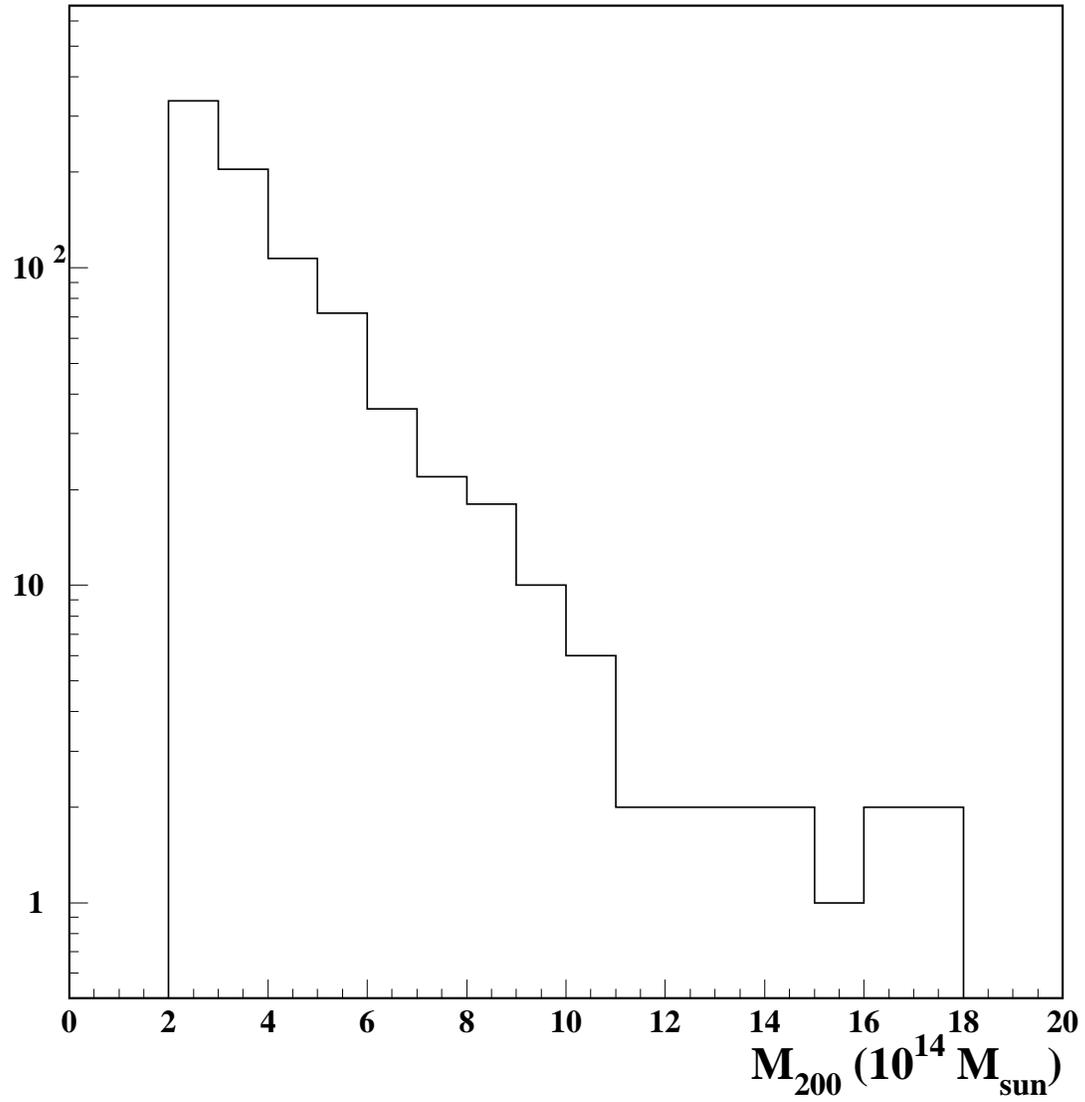}
\end{center}
\caption{The mass distribution for our sample of inspected clusters.
The masses were estimated from the $\sigma_v$-$N_g$
relation mentioned in the text and from Eqn.~(\ref{mass}).}
\label{massDist}
\end{figure}
\clearpage

\section{Search by Visual Inspection}

Previous arc surveys have generally used visual inspection to locate arc
candidates, which are then usually followed up with deeper imaging
in better seeing. 

The SDSS images of the 825 selected clusters were visually
inspected for arcs. For each cluster, we used a SDSS
coaddition code to build an image covering an angle
corresponding to a physical scale of  1 Mpc (assuming $H_0 = 100$
km/s, $\Omega_m = 0.3$, and $\Omega_\Lambda = 0.7$). The
inspector was then presented with 4 simultaneous images
using the {\em ds9} image display program: grayscale
images of the $g$, $r$, and $i$ bands and a color image that
combines the three bands. The grayscale images were displayed using
the {\em ds9} ``zscale'' algorithm, while the color image
was displayed using a fixed surface brightness range (the same
range was used for all clusters).  A single author
of this work (T.L.) scanned the 825 images, while three
other authors inspected images of the 300 highest $N_g$ clusters in
the sample. Candidates were recorded for later consideration
and for potential follow-up observations (\S\ref{follow-up}).

\subsection{Visual Inspection Selection Function}

\begin{figure}
\begin{center}
\plotone{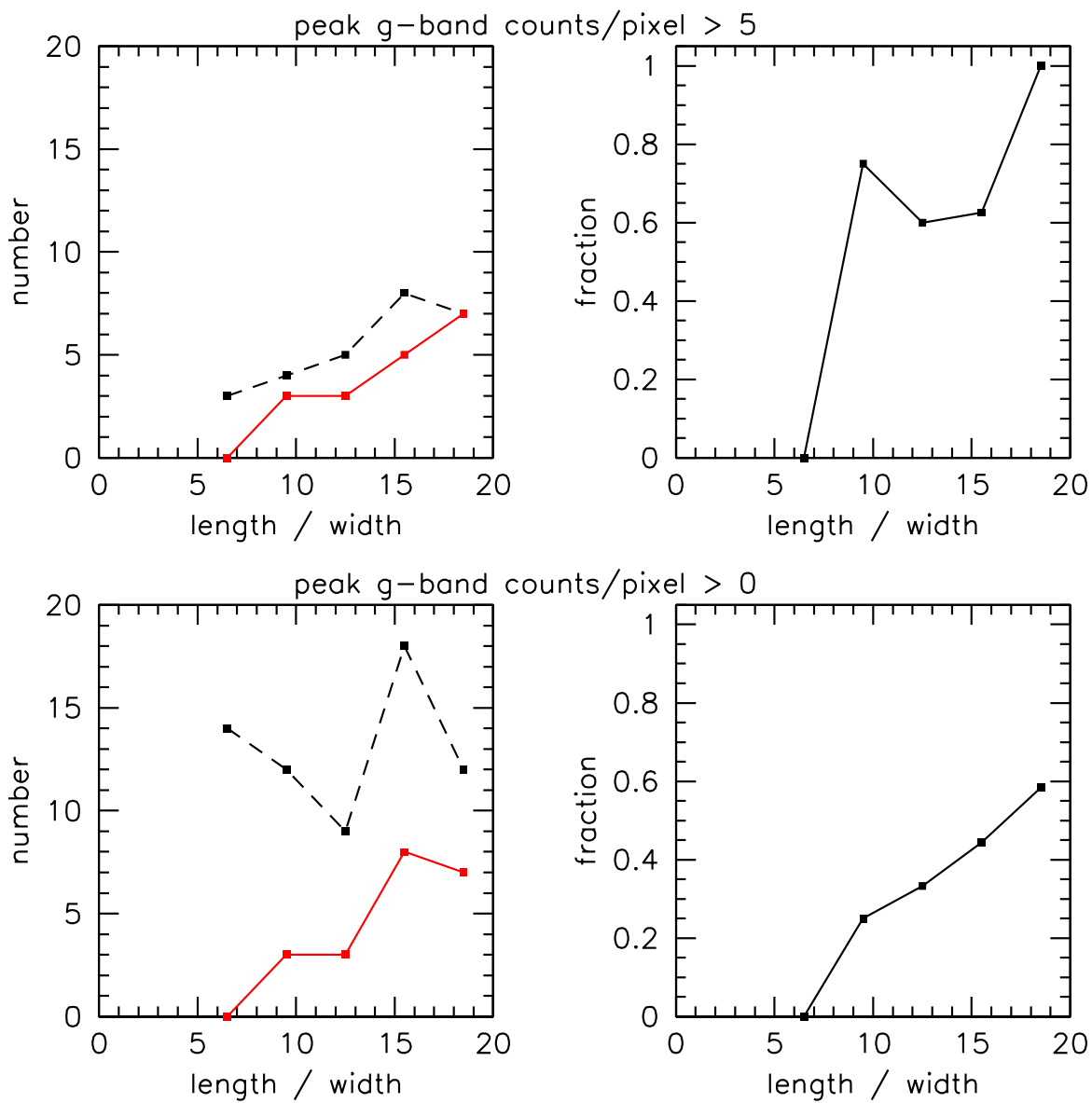}
\end{center}
\caption{Efficency for finding simulated arcs as a function of the
length-to-width ratio $l/w$. 
The panels on the left show the number of simulated arcs included
in the images (dashed black) and the number of these arcs recovered by the
visual inspection (solid red).  The panels on the right show the
efficiency for recovering the simulated arcs.  The top
panel shows the results for arcs with more than 5
counts/pixel in peak surface brightness 
(the conversion of counts per pixel to surface brightness
is shown in Fig.~\ref{eff-g}) and the bottom panel
shows the efficiency for the whole sample. } \label{eff-lw}
\end{figure}

\begin{figure}
\begin{center}
\plotone{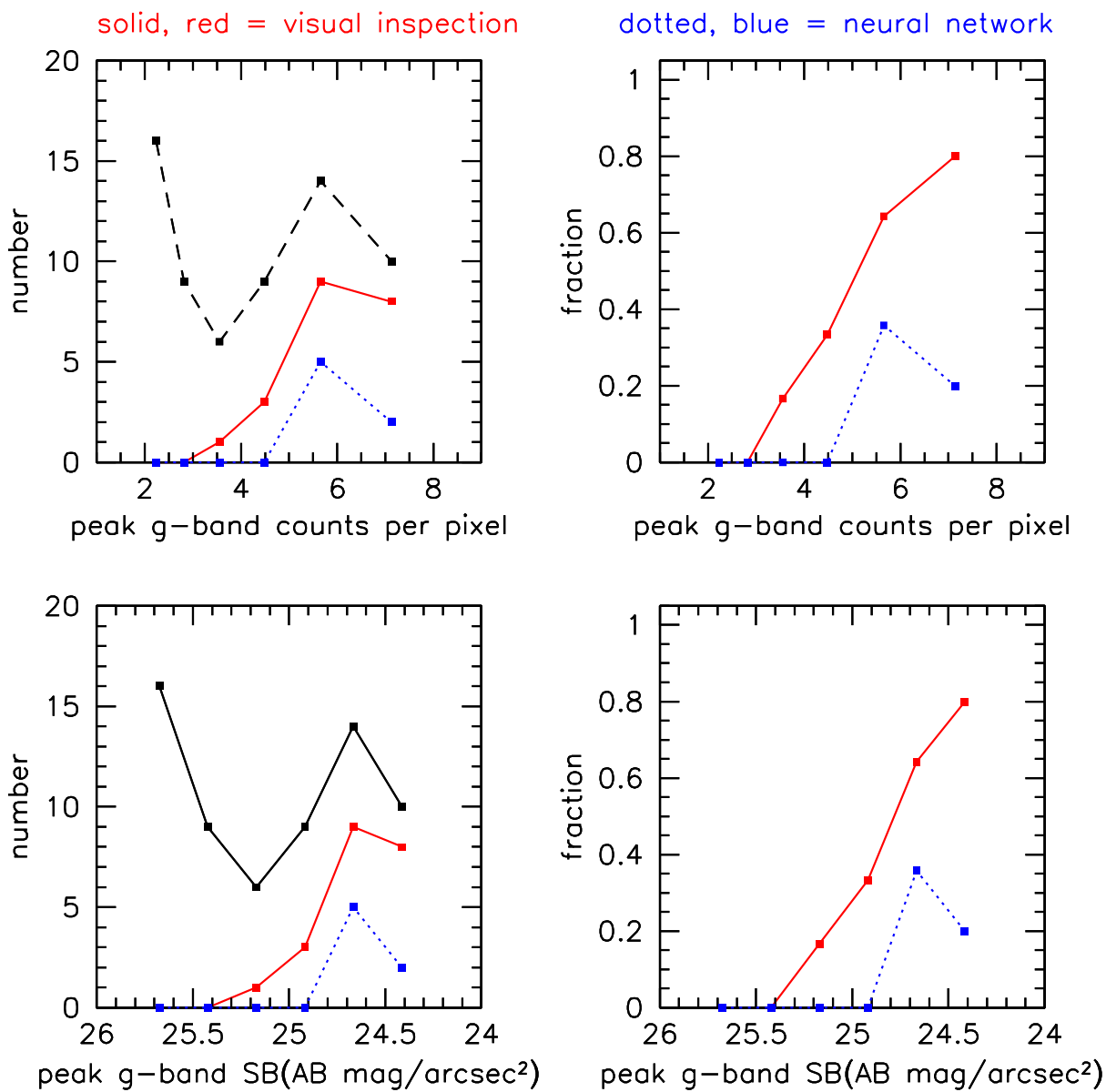}
\end{center}
\caption{Efficiency for finding simulated arcs as a function of
$g$-band peak surface brightness, in units of either
counts per pixel (top panels) or AB magnitudes per arcsec$^2$ 
(bottom panels).  The panels on the left show the
number of simulated arcs included in the images (dashed black), the number 
recovered by the visual inspection (solid red), and the number
recovered by the automated search (dotted blue). The panels on the right
show the efficiency for recovering the simulated arcs.}
\label{eff-g}
\end{figure}

A selection function is a prerequisite for the proper use of
arc statistics to constrain cosmology. In order
to calculate a selection function for our arc
search, simulated arcs were added to a fraction of the images in our
cluster sample. These objects were included before the images were
inspected for arcs and without the knowledge of the scanner doing
the visual inspection.  Each added arc consisted of a section of a circle
centered on the brightest cluster galaxy identified by the
maxBCG algorithm.  We added the arcs using
uniform distributions in log peak surface brightness and 
in length-to-width ratio ($l/w = 5-20$) in order to derive the
arc detection efficiency as a function of these arc parameters.
The simulated arcs have a flat spectrum in AB magnitudes, 
similar to typical observed gravitational arcs from other surveys
(this makes the simulated arcs much bluer than the cluster galaxies).
The typical length
for the simulated arcs was $l \approx 10''$, and they were inserted
in the images at $5''-45''$ from the brightest cluster 
galaxy (BCG).  The simulated arcs have a surface brightness 
profile that is uniform along the tangential direction 
and is a gaussian along the radial
direction, with a FWHM = $1.2''$ ($\approx w$), similar to the 
typical SDSS seeing. The selection functions obtained with this
technique are presented in Figs.~\ref{eff-lw} and 
\ref{eff-g}.

\clearpage

\clearpage

\clearpage

For the visual inspection, the detection efficiency reaches 50\% at 
about 24.8 mag/arcsec$^2$ in the peak g-band surface brightness
and for $l/w \approx 10$. 
The primary causes for missed arcs are low surface brightness
and small length-to-width ratio.
However, the recovery efficiency does not reach 100\% even for 
high $l/w$ ratio, high surface brightness arcs, due to
blending problems with bright galaxies and stars on the images. 
Note that the measurement of our selection function is somewhat 
limited by small-number statistics,
because we decided not to populate a significant fraction of our
images with simulated arcs (we added 65 simulated arcs in total,
i.e., to about 8\% of the 825 inspected clusters).

\section{Search by Algorithm}

While sophisticated arc-finding algorithms have been published previously
\citep[e.g.,][]{arc_alg,arc_alg2,alg_alard,alg_Starck,siedel06}
we are not aware of an arc search that
has been performed in a fully automated way.
Clearly if we are to examine $10^5$ clusters we will need to
use an algorithm (as opposed to a human eye inspection). We
have developed such an algorithm, based on ideas taken from related
techniques  used in experimental particle physics for tracking
particles in cloud and bubble chambers. In this work we will
set a benchmark comparison between a simple algorithm for arc
detection and a visual inspection.

\subsection{The Algorithm}

Our algorithm is composed of three steps: image preparation,
candidate pre-selection, and final arc selection.

Step 1 is image preparation, where we scale the $g,r,i$ images by 
the variances of
their respective sky backgrounds, normalize the variances to 1.0, 
average the three images together, and then run SExtractor \citep{sextractor} 
to produce an object image.  This step is intended to
enhance the signal-to-background ratio of the arc-like objects.
The averaging of the variance-normalized images
follows the ideas of \cite{szalay}, and the intent is to produce an image
where the pixel values are a $\chi^2$ of the hypothesis that the pixel
consists of only sky.
The SExtractor-produced object image has eliminated isolated pixels above
threshold (due to  noise) that do not belong to any object, retaining
only pixels belonging to objects dectected above threshold.
We perform a deblending of the SExtractor objects by 
separating each object into sub-objects which correspond
to groups of contiguous pixels above threshold, and those sub-objects 
that are separated from another sub-object by fewer than 2 pixels are 
merged into one object.

Step 2 is candidate pre-selection, where the basic shape of the
candidates is characterized by measuring the major and minor 
axes of each object, and where objects that show elongated shapes are kept
for further study. The major axis ($r_1$) is defined as the largest
distance between any pair of the $N$ pixels belonging to the object,
and the minor axis  ($r_2$) is defined as the maximum distance
between pixels perpendicular to the main axis. Elongated
objects, where $r_1/r_2 \ge 1.4$, are chosen for further analysis.
Step 2 simply reduces the number of objects to be studied, in order
to reduce computation time per image. It is designed to be both
complete and efficient, keeping $\approx 100\% $ of the arc-like objects
while providing about a factor of 100 in rejection against the rest 
of the objects detected by SExtractor.

Step 3 is final arc selection, where the remaining objects have their
radius of curvature measured.  Those radii, plus other parameters measured 
in step 2, become the input for a neural net (NN) trained to select 
simulated arcs. The output of this NN is used to determine if an object is
a good arc candidate.
The radius of curvature ($R$) is measured using a least squares fit.
The selection of good arc candidates then uses the following quantities:
\begin{itemize}
\item $r_1/(2 \pi R)$:
    the fraction of the circumference covered by the arc.
    This is typically $>0.2$ for objects with significant curvature.
\item $N / (\pi R^2)$:
    the fraction of the circle area covered by pixels above threshold.
    This is typically $<0.15$ for good candidates.
\item $\chi^2/N$: the goodness of fit.
\item $r_1$: the length of the long axis.
    This ensures significant size for the candidates, $>20$ pixels
    (1 pixel $= 0.396''$).
\end{itemize}
The 4 quantitites described above are then presented to a NN with 8
nodes in a hidden layer and the output is trained (by back propagation) to
identify simulated arcs. The NN was trained using a separate
sample of 100 bright simulated arcs, where the peak surface brightness 
was fixed at 20 counts per pixel in the $g$ band.

\subsection{Algorithm Selection Function}

The efficiency obtained with this algorithm is somewhat lower than
that seen for the human scan and is shown in Fig. \ref{eff-g}. The
detection efficiency reaches 40\% at 24.7 mag/arcsec$^2$ in the $g$ band
and for $l/w \approx 10$; the curve is consistent with the eye scan
curve shifted to a brighter surface brightness by $\approx 0.4$ mags. 
More development will be needed to bring this algorithm to the level of
efficiency of the visual inspection. However, this analysis
provides a quantitative benchmark for the comparison of a human search 
with a simple algorithm search applied to the same cluster sample.

\section{Candidate Follow-up} \label{follow-up}

%
%

The six best candidates, all from the visual inspection, were
selected for follow-up. These arc
candidates were observed using the SPIcam imager on the Apache Point
Observatory 3.5-meter telescope.  Images were takin in the SDSS
$g$, $r$, and $i$ filters. 
The observations were done the night of April 29-30, 2006, during dark time.
Table~\ref{tab-followup} shows the amount of time spent observing
each object with the three filters. The images were reduced using
standard bias subtraction and flat-fielding techniques.
The images were also registered and coadded. Typical seeing on
the individual images ranged from $1.1''$ to $1.5''$.

The candidates were inspected by the authors and discarded based on
judgements that they were most likely ``junk'' or edge-on galaxies.
Room for improvement in this system exists and we plan on putting
into place a set of criteria for systematic follow-up. The criteria
useful for SDSS data are obvious from an examination of Fig.
\ref{RCS_arc}: $l/w$, surface brightness, radius of curvature, and
color.

\begin{deluxetable}{lcccccc}
\tablecolumns{7}
\tablewidth{0pc}
\tablecaption{Candidate Observations\label{tab-followup}}
\tablehead{
    \colhead{Name} & \colhead{RA}   & \colhead{Dec}    & \colhead{z} &
    \colhead{g (min)} & \colhead{r (min)}    & \colhead{i (min)}}
\startdata
SDSS+139.5+51.7+0.24   & 09:17.59.5 & 51:42:06 & 0.24 & 15 & 5 & 5 \\
SDSS+184.4+3.6+0.08   & 12:17:28.0 & 03:36:17 & 0.08 & 15 & 5 & 5 \\
SDSS+117.7+17.7+0.19   & 07:50:48.9 & 17:40:43 & 0.19 & 15 & 5 & 5 \\
SDSS+175.3+5.8+0.12 & 11:41:13.5 & 05:48:28 & 0.12 & 15 & 5 & 5 \\
SDSS+205.5+26.4+0.10  & 13:41:49.1 & 26:22:25 & 0.10 & 15 & 5 & 5 \\
SDSS+139.1+5.9+0.14 & 09:16:21.3 & 05:53:18 & 0.14 & 15 & 5 & 5 \\
\enddata
\tablecomments{SPIcam observations.  Exposure times are given in minutes.
The redshifts quoted for the clusters are photometric redshifts
from the maxBCG algorithm and have a dispersion of $\sigma_z = 0.015$.}
\end{deluxetable}

\section{Results of the Searches}

Our survey of
825 SDSS galaxy clusters resulted in no convincing candidates for
giant gravitational arcs.

\begin{figure}
\begin{center}
\plotone{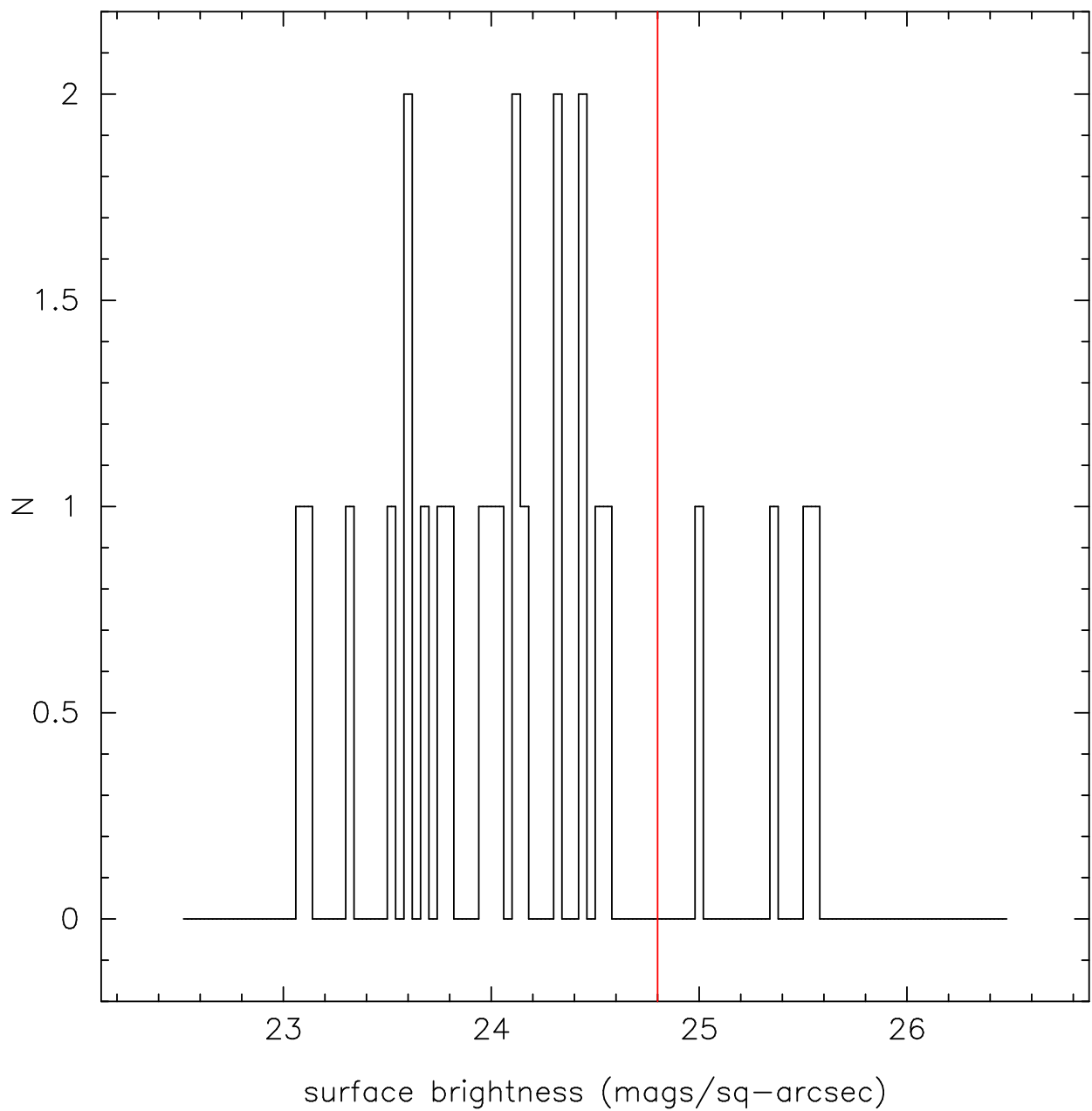}
\end{center}
\caption{The surface brightnesses of the arcs found in previous
ground based searches. The surface brightness is reported in the AB
system; given the simplifying assumption of a
flat spectrum arc, we need not report the bandpass, though
the data were in fact observed through a variety of filters
($B,V,R,I$). The vertical red line is the $50\%$ efficiency limit of our
visual search selection function.
} \label{sb}
\end{figure}

The selection function we have computed shows that we are $\ge 50\%$
efficient at SB $\le 24.8$ mags/arcsec$^2$ and for $l/w \ge 10$. The
latter, given our mean seeing of $1.4''$, corresponds to arcs longer
than $14''$. Fig.~\ref{sb} shows the surface brightness distribution
of the arcs found by \citet[in seeing of $\sim 1.0$'']{petrosian}, 
by \citet[$\sim0.90''$, their Table 2]{luppino},
by \citet[$\sim0.55''$]{zaritsky03}, and by \citet[$\sim0.85''$]{RCS}.
These are all the published
ground based surveys, and no attempt to homogenize the
seeing has been made. The surface brightnesses have been converted
into the AB system from the Vega system. We assume that the arcs have
flat spectra so that no change is necessary for the different
bandpasses. 
We see from Figure~\ref{sb} that our surface brightness limit
is in fact exceeded by the majority of these arcs.  
The observed surface brightness
for arcs that are unresolved in width is a function of the seeing
obtained, and the data here include a variety of seeing.  
However, all are better than our sample's mean seeing of 
$1.4''$, so if this plot were homogenized to the SDSS seeing most 
of these arcs would fall below our surface brightness limit. Seeing is 
thus the dominant limiting factor in our efficiency, and
it is the relatively large PSF of the SDSS data that 
limits our detection to high surface brightness,
quite long arcs with $l/w \ge 10$.
On the other hand, the arcs that we can in fact find in the SDSS
will therefore also be particularly interesting objects for follow-up 
observations.

\clearpage
\clearpage

\subsection{Upper Limits on the Observable Arc Production}

We use our search results to establish upper limits on the production of
observable giant arcs by galaxy clusters. Assuming Poisson
statistics, an observation of 0 arcs for a given inspected cluster
indicates that the 95\% C.L.\ upper limit on the expected number of 
observable arcs per cluster is 3 (i.e., the probability of observing
0 events for a Poisson distribution with mean 3 is 
$3^0 e^{-3} / 0! \approx 0.05 = 1-0.95)$. 
Then, given a null result after inspecting $N_i$ clusters in a redshift
bin $i$, the 95\% C.L.\ upper limit on the expected number
of observable arcs per cluster, $\Sigma^{(i)}_{95\%}$, may
be determined from the product of the individual Poisson probabilities
using the relation
\begin{equation}
[ \ (\Sigma^{(i)}_{95\%})^0 \ e^{-\Sigma^{(i)}_{95\%}} \ / \ 0! \ ]^{N_i} = 
e^{-\Sigma^{(i)}_{95\%} N_i} = 0.05 \ ,
\end{equation}
from which we find that
\begin{equation}
\Sigma^{(i)}_{95\%} = \frac{ 3 } { N_i} . \label{eq:limit}
\end{equation}

\begin{figure}
\begin{center}
\plotone{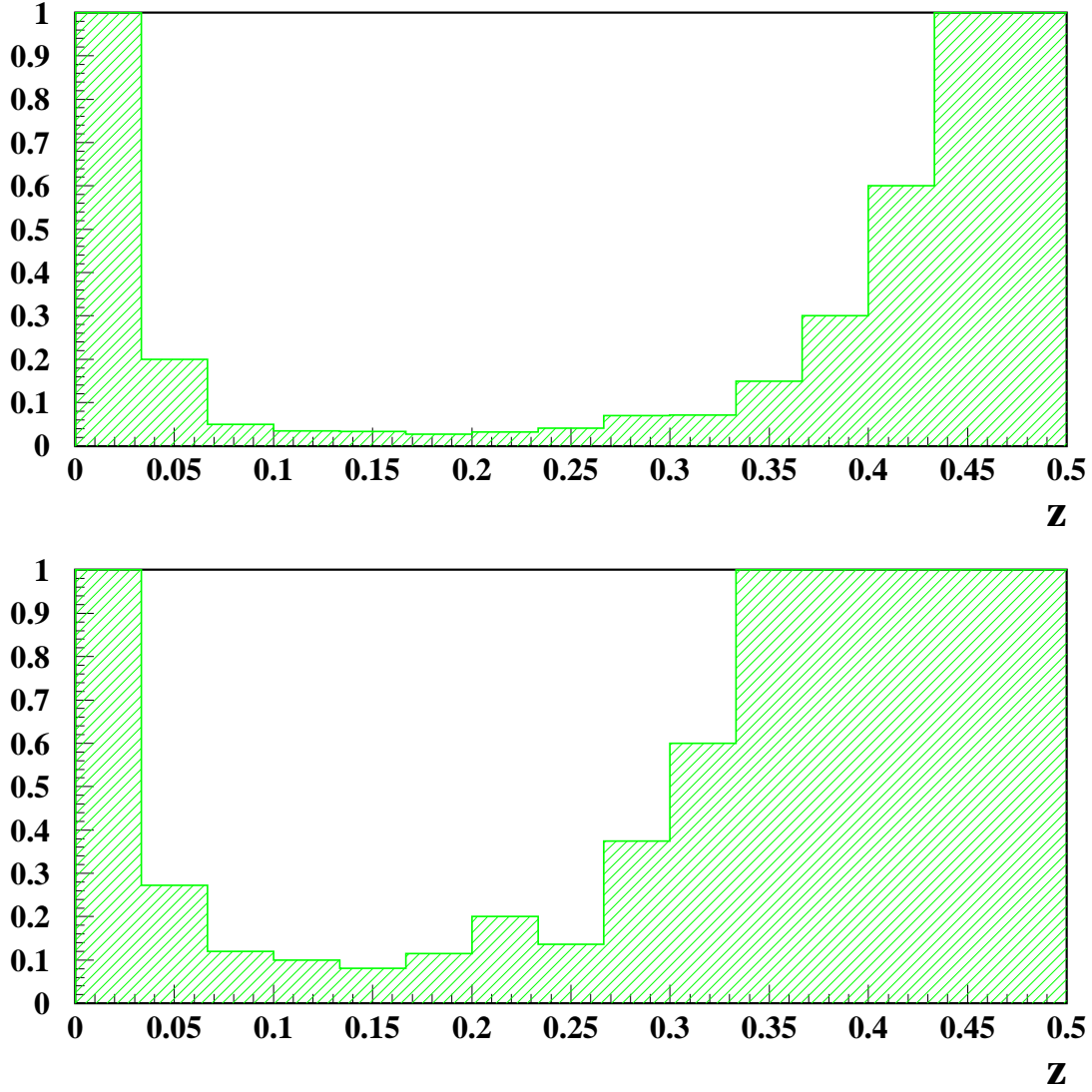}
\end{center}
\caption{ The 95\% C.L.\ upper limit for the probability for a
cluster to produce a gravitational arc visible in SDSS imaging,
assuming Poisson statistics in our sample of 825 clusters. The limit
is shown as a function of redshift for $M_{200} = 2 - 4 \times
10^{14} M_\sun$ clusters in the top panel and 
for $ M_{200} > 4 \times 10^{14} M_{\sun}$ clusters in the bottom panel.}
\label{limit}
\end{figure}

$\Sigma^{(i)}_{95\%}$ is shown in Fig.~\ref{limit}
for clusters with estimated masses $M_{200} > 2 \times 10^{14} M_{\sun}$
and $M_{200} > 4 \times 10^{14} M_{\sun}$.
It should be noted that our upper limits correspond to arcs
observable by the SDSS according to the efficiency for detection shown
in Figs.~\ref{eff-lw} and \ref{eff-g}. The lower
panel in Fig.~\ref{limit} corresponds to the more massive clusters
($M_{200} > 4 \times 10^{14} M_{\sun}$) in our sample, and from the
plot it is clear that we only have the
statistics for a significant limit in the range $0.1 < z < 0.3$.

\clearpage
\clearpage

\subsection{Comparison With Previous Works}

In order to understand the compatibility of our results with
previous arc searches, we perform a direct  comparison with these
searches over similar redshifts. The cluster 
selection function is a considerable source of uncertainty 
which is hard to quantify for this comparison, ans so to reduce the 
impact of this effect we have selected on each
sample clusters with similar mass estimators.
The sample of X-ray luminous galaxy
clusters constructed by \cite{smith_HST} is a good place to start.
This sample has a total of 10 rich clusters with $M_{200} > 4 \times
10^{14} M_{\sun}$ 
at $0.17 \le z \le 0.26$, where the mass is estimated from X-ray
luminosities \citep{massLx}. The sample was imaged with HST and the
geometrical properties of the arcs were published by
\cite{HST_sands}. In order to estimate the expected number of arcs
observable in the SDSS, the parameters of the arcs are degraded from
those observed in HST imaging (with a typical PSF of 0.15'') to the
image quality of the SDSS (with a typical PSF of 1.4''). This conversion
is done assuming that the arc is well resolved in HST imaging, i.e.,
that the width is entirely due to the object, not the PSF. The width
is then increased by adding in quadrature the average seeing for
the SDSS, effectively reducing the $l/w$ ratio. The increased
width results in an increased area for the object, which translates
into a reduction in surface brightness. Of the arc candidates observed
with HST in these clusters, only 3 remain with $l/w > 10$
after degradation to SDSS seeing. Of the surviving arcs only 1 
has a surface brightness inside our detection limit.
Thus the sample of \cite{smith_HST} gives a 10\%
probability of having an arc observable by SDSS for clusters with
$M_{200} > 4 \times 10^{14} M_{\sun}$ and $0.15 < z < 0.3$. 
This result lies inside the 95\% C.L. presented in Fig. \ref{limit}.

A similar comparison can be done with the \citet{luppino} sample of
38 EMSS X-ray selected clusters
with $L_x \ge 2\times10^{44}$ ergs/sec at $0.15 \le z \le 0.82$. As
before, the parameters for the arcs were transformed into expected
parameters for the SDSS using the seeing applicable
to each arc observation carried out by \citet{luppino}. After this procedure, 
no arc from this sample is expected to be observable in SDSS imaging. 
Again, this result is consistent with the upper limits derived from
our analysis.

The results obtained in this work are also consistent with
a search for giant gravitational arcs in the RCS \citep{RCS}. 
The RCS cluster sample covers a redshift range $0.35 < z < 0.95$,
and after visual inspection of approximately 900
clusters with $M_{200} > 4 \times 10^{14} M_{\sun}$,
3 strong lensing systems with giant arcs were found, all of them with 
$ z > 0.60 $. The result from the RCS indicates a smaller
probability for a low redshift
cluster to produce an arc, compared with a cluster of
similar mass at higher redshifts. The bulk of the 
clusters used for the arc search in the present work have 
redshifts $0.1 < z < 0.3$, lower than the arc clusters found by the RCS,
and so our null result is also consistent with the RCS result
that low-redshift clusters are less efficient for the production of arcs.

We are aware of 4 giant arc systems recovered or discovered 
in single-pass SDSS imaging: one found by the RCS \citep[][see below]{RCS}, 
one found by \citet[][RX J1133]{sand04},
a new system described in the Appendix, and another new system
that is described in \citet{sahar_arcs}. 
Three of the arc systems are lensed by clusters
with $N_g \sim 10-20$, beneath the threshold of our current work,
and these also have $0.35 \le z \le 0.45$, beyond our catalog redshift
limit though well within the reach of SDSS cluster finders. The
fourth is RCS1419.2+5326 at $z=0.64$: our cluster finders can discover
objects with these properties, but with
$N_g \sim 5$ (as observed by the SDSS; the real $N_g$ is
much higher) the catalogs are far from
the $90\%$ efficiency level of the $z < 0.3$ catalogs
(the actual efficiency may be of order $1-5\%$). 
It is worth noting that the two high
surface brightness arcs found by \cite{petrosian} are just south of
the limits of the SDSS Stripe 82 region. Had both clusters been 1 degree
farther north, they would likely have been found in our survey.


\section{Summary}

The potential for observing arcs in SDSS imaging has been
demonstrated by the observation of known arc systems in the
multiband imaging of the survey (see Fig.~\ref{RCS_arc}) and by the
discovery of new arcs \citep[this paper and][]{sahar_arcs}. 
In this work, we presented
a systematic search for gravitational arcs in a sample of SDSS
clusters. A total 825 clusters with $N_g>20$ 
(corresponding to $M \gtrsim 2 \times 10^{14} M_{\sun}$) 
have been visually inspected for arcs and have also been
processed with an automated arc search algorithm. The efficiency of 
arc detection for both techniques has been quantified by
using simulated arcs added to the inspected images. The results
indicate a significantly higher efficiency for the eye
scan and point to the need for developing a more efficient
algorithm for the automated search. No gravitational
arcs were seen in the sample of clusters analyzed. This result is
consistent with previous surveys done over a similar redshift
range ($0.1 < z < 0.3$).

\acknowledgments
\section{Acknowledgements}

Funding for the SDSS and SDSS-II has been provided by the Alfred P. Sloan Foundation, the Participating Institutions, the National Science Foundation, the U.S. Department of Energy, the National Aeronautics and Space Administration, the Japanese Monbukagakusho, the Max Planck Society, and the Higher Education Funding Council for England. The SDSS Web Site is http://www.sdss.org/.

The SDSS is managed by the Astrophysical Research Consortium for the Participating Institutions. The Participating Institutions are the American Museum of Natural History, Astrophysical Institute Potsdam, University of Basel, Cambridge University, Case Western Reserve University, University of Chicago, Drexel University, Fermilab, the Institute for Advanced Study, the Japan Participation Group, Johns Hopkins University, the Joint Institute for Nuclear Astrophysics, the Kavli Institute for Particle Astrophysics and Cosmology, the Korean Scientist Group, the Chinese Academy of Sciences (LAMOST), Los Alamos National Laboratory, the Max-Planck-Institute for Astronomy (MPIA), the Max-Planck-Institute for Astrophysics (MPA), New Mexico State University, Ohio State University, University of Pittsburgh, University of Portsmouth, Princeton University, the United States Naval Observatory, and the University of Washington.

\appendix
\section{Appendix: Hall's Arc} \label{appendix}

We report here an arc discovered serendiptously by one of us (P.B.H)
in 2004. It is widely known inside the SDSS Collaboration but has
not been published previously. The purely serendipitous discovery
was made during inspection of the images of spectra classified
UNKNOWN (the initial reduction of the SDSS spectrum had bad
spectrophotometry and could not be classified). There is a cluster
clearly visible in the images: the BCG has a SDSS spectrum and is at
$z=0.44$, and it is coincident with NVSS J014656-092952. The SDSS
spectrum shows two objects superimposed, the $z=0.44$ galaxy and a
star -- the latter is what the SDSS pipeline reports. 

We ran the maxBCG code at the position of the BCG and measured $N_g
= 12$, but this is a lower limit. The cluster is above the redshift
where the $N_g$ is complete -- at this redshift two effects
compromise the $N_g$ measurement:
(1) $0.4L^\star$ is fainter than the useful limiting magnitude of the
SDSS, and (2) the colors of objects near $0.4L^\star$ become noisy
enough to scatter outside the color box used to determine cluster
membership. A correction must be applied to $N_g$ for these effects.
As we are not attempting to use this arc in our analysis we defer
the measurement of the corrected $N_g$ to a later paper. We have
tabulated our information about the cluster in
Table~\ref{tab-hall-cluster}.

There are three arcs. The longest one was split by the SDSS photometry pipeline
Photo into two objects, a and b. We report on the object parameters in
Table~\ref{tab-hall}. Three of the four objects have photometric
redshifts of about 0.6-0.7; the fourth is about $3\sigma$ away
(though we note that for faint objects $r > 20$
the SDSS photometric redshifts are less reliable because the 
spectroscopic training sets available are less extensive than at 
brighter magnitudes).
We measured the parameters of the three arcs in the SDSS $g$-band image. These
are listed in Table~\ref{tab-hall2}. It is likely that we would
have found this arc in our survey (if it had been in our sample catalog)
because it is remarkably clean:
there are no stars or cluster galaxies that are projected
onto the images of the arcs.
Deeper imaging data for this system may be found in \cite{Hennawi06}.

\begin{deluxetable}{lcccc}
\tablecolumns{5}
\tablewidth{0pc}
\tablecaption{Hall's Arc: the Cluster\label{tab-hall-cluster}}
\tablehead{
    \colhead{Name} & \colhead{RA}   & \colhead{Dec}    & \colhead{$z$} & \colhead{$N_g$}}
\startdata
SDSS+26.733-9.497+0.44 & 01:46:56.00 & -09:29:52.4 & 0.44 & $\ge12$\tablenotemark{a}
\enddata
\tablenotetext{a}{\ Lower limit; see text.}
\end{deluxetable}

\tabletypesize{\small}
\begin{deluxetable}{lcccccccc}
\tablecolumns{9}
\tablewidth{0pc}
\tablecaption{Hall's Arc: Positions, Magnitudes, and Photo-z's\label{tab-hall}}
\tablehead{ \colhead{Arc}
    & \colhead{RA} & \colhead{Dec}
    & \colhead{$g$}
    & \colhead{$u-g$} & \colhead{$g-r$}
    & \colhead{$r-i$} & \colhead{$i-z$}
    & \colhead{$z_{phot}$\tablenotemark{a}}
    }
\startdata
1a & 26.73210 & -9.50009 &
    $21.85\pm0.15$ & $1.3\pm1.4$
    & $0.57\pm0.21$ & $0.68\pm0.19$ & $0.09\pm0.51$ & $0.61\pm0.14$\\
1b & 26.73051 & -9.49967 &
    $21.81\pm0.21$ & $0.39\pm0.92$
    & $0.75\pm0.27$ & $0.63\pm0.22$ & $-1.8\pm2.6$ & $0.57\pm0.23$ \\
2 & 26.73063 & -9.49525 &
    $21.66\pm 0.10$ & $0.45\pm0.46$
    & $0.53\pm0.14$ & $0.73\pm0.12$ & $0.29\pm0.28$ & $0.67\pm0.08$ \\
3 & 26.73641 & -9.49673 &
    $21.77\pm 0.15$ & $0.66\pm 0.81$
    & $0.03\pm 0.27$ & $0.15\pm 0.39$ & $-1.1\pm 2.1$ & $0.15\pm 0.15$\\
\enddata
\tablecomments{SDSS data. We retain the SDSS decimal degrees convention for RA and Dec. The magnitudes and colors are SDSS model magnitudes and are
corrected for extinction \citep[see][and references therein]{DR5}.}
\tablenotetext{a}{The mean of the SDSS DR5 photoz and photoz2 \citep{DR5}
photometric redshifts and errors.}
\end{deluxetable}

\tabletypesize{\normalsize}
\begin{deluxetable}{cccc}
\tablecolumns{4}
\tablewidth{0pc}
\tablecaption{Hall's Arc: Arc Parameters\label{tab-hall2}}
\tablehead{ \colhead{Arc}
    & \colhead{$l$} & \colhead{$l/w$}
    & \colhead{mean SB (mag/arcsec$^2$)}
    }
\startdata
1 &$17.2''$ &11.2 & 24.5   \\
2 &$10.0''$ &6.5 &24.6    \\
3 &$3.8''$ &2.5  &23.7   \\
\enddata
\tablecomments{SDSS $g$-band measurements. The $g$-band PSF is $1.53''$.}
\end{deluxetable}

\begin{longtable}[!hb]{lrrcccr}
\hline
Name & RA & DEC & $z$  & $ N_{g} $ & $M_{200}$   \\
\hline
      SDSS+239.6+27.2+0.11  &  239.634     &  27.1808     &  0.109     &  94          &  31.2                      \\
      SDSS+230.6+27.7+0.08 &  230.6       &  27.7144     &  0.083     &  88          &  28.1                      \\
     SDSS+213.6-00.3+0.13  &  213.618     & -0.3301      &  0.133      &  69          &  17.7                      \\
      SDSS+117.7+17.7+0.19 &  117.721     &  17.6768     &  0.192     &  68          &  16.7                      \\
      SDSS+258.2+64.0+0.08 &  258.227     &  63.9924     &  0.081     &  67          &  17.2                      \\
      SDSS+227.8+05.8+0.08 &  227.75      &  5.7828      &  0.081     &  65          &  16.3                      \\
      SDSS+197.9-01.3+0.20 &  197.886     & -1.3329      &  0.203     &  65          &  15.3                      \\
      SDSS+234.9+34.4+0.25  &  234.893     &  34.4369     &  0.249     &  63          &  14.1                      \\
      SDSS+250.1+46.7+0.25 &  250.098     &  46.7028     &  0.242      &  62          &  13.7                      \\
     SDSS+110.4+36.7+0.15 &  110.36      &  36.7383     &  0.148     &  62          &  14.4                      \\
     SDSS+227.6+33.5+0.12  &  227.584     &  33.486      &  0.116     &  59          &  13.4                      \\
     SDSS+250.8+13.4+0.20 &  250.843     &  13.363      &  0.199      &  58          &  12.5                      \\
     SDSS+126.3+47.1+0.13 &  126.32      &  47.1424     &  0.129     &  58          &  12.9                      \\
     SDSS+203.8+41.0+0.26 &  203.818     &  41.0132     &  0.264       &  55          &  10.9                      \\
     SDSS+139.5+51.7+0.24  &  139.481     &  51.7163     &  0.236     &  54          &  10.7                      \\
     SDSS+255.7+34.1+0.11 &  255.69      &  34.0611     &  0.109     &  54          &  11.5                      \\
     SDSS+228.8+04.4+0.11 &  228.823     &  4.397       &  0.107     &  53          &  11.1                      \\
     SDSS+137.3+11.0+0.18 &  137.329     &  10.9857     &  0.179     &  53          &  10.7                      \\
     SDSS+184.4+03.6+0.08 &  184.381     &  3.6157      &  0.089     &  52          &  10.9                      \\
     SDSS+216.5+37.8+0.16  &  216.475     &  37.7915     &  0.164     &  50          &  9.7                       \\
\hline
\caption{List of the most 20 most massive inspected clusters in our sample. The full
sample is available electronically.} 
\label{clusterlist}
\end{longtable}


\end{document}